\documentclass[prb,twocolumn,aps,showpacs,preprintnumbers]{revtex4}
\usepackage{graphicx}
\usepackage{dcolumn}
\usepackage{bm}

\begin{document}

\preprint{Draft}

\title{Spin Hall effect, Hall effect and spin precession in diffusive normal metals}

\author{R.V. Shchelushkin and Arne Brataas}
\affiliation{Department of Physics, Norwegian University of Science 
and Technology, N-7491 Trondheim, Norway}
\date{\today}

\begin{abstract}
We study transport in normal metals in an external magnetic field. This system exhibits an interplay between a transverse spin imbalance (spin Hall effect) caused by the spin-orbit interaction, a Hall effect via the Lorentz force, and spin precession due to the Zeeman effect. Diffusion equations for spin and charge flow are derived. The spin and charge accumulations are computed numerically in experimentally relevant thin film geometries. The out-of-plane spin Hall potential is suppressed when the Larmor frequency is larger than the spin-flip scattering rate. The in-plane spin Hall potential vanishes at zero magnetic field and attains its maximum at a finite magnetic field before spin precession starts to dominate. Spin-injection via ferromagnetic contacts creates a transverse charge Hall effect that decays in a finite magnetic field due to spin precession. 
\end{abstract}


\maketitle

Spintronics is a new subfield of research which could provide a new class of low power and high speed electronic devices.
This requires an understanding of spin-injection, spin manipulation and spin detection. In many metals, spins are affected by the spin-orbit interaction which is often 
considered a nuisance causing a decay of the injected spin flow. However, the spin-orbit interaction can also be used to manipulate the spins in a desired way and even to cause 
a finite spin polarization of an initially unpolarized electron flux. This latter spin Hall effect \cite{dyak1,dyak2,chaz,hirsh,Exper,murakami,sharma,wolf,ganichev,sinitsyn,culcer,burkov,new1} has recently attracted considerable interest.

Spins are influenced by a magnetic field. A magnetic field governs the diffusion of particles via spin precession induced by the Zeeman effect \cite{Jedema} and via the Lorentz force. This alters the spin and charge currents and the spin and charge accumulations. There are already some theoretical works on the influence of an external magnetic field on the spin Hall effect. Refs.\ \onlinecite{chang,shen} discuss the influence of a magnetic field on the intrinsic \cite{murakami,sinova} spin Hall effect. On the other hand, within our knowledge, theoretical studies of the extrinsic spin Hall effect in normal metals are limited to the regime of zero magnetic field.  

Recently, the first clear optical observation of the spin Hall effect in semiconductors was presented \cite{Exper}. In this experiment, the dominant contribution to the spin Hall effect was extrinsic \cite{dyak,khae,zhang} and caused by spin-orbit scattering at impurities. It was also shown that an applied in-plane magnetic field causes precession and suppression of the spin accumulation via the Hanle effect.\cite{hanle}

We address how a magnetic field affects the extrinsic spin Hall and Hall effects in {\it normal metals}, where spin-injection and spin-detection can be done by electrical contact.\cite{Jedema} To this end, we study the extrinsic spin Hall effect in normal metals in a external magnetic field. We derive the diffusion equations for spin and charge flow and calculate the spin Hall and also Hall voltage as functions of magnetic field and sample geometry. We consider a normal metal thin film with weak extrinsic spin-orbit interaction.  The computed magnetic field dependence can assist in the understanding of future all-electrical experimental detection of the spin Hall effect in normal metals.
 
Our transport theory is based on the Keldysh formalism \cite{schwab,rammer2}. We first employ the quasiclassical approximation since the system size is longer than the Fermi wavelength and second use the diffusion approximation valid when the system size is larger than the mean free path. By computing the Keldysh component, we derive finally the spin and charge current and density distribution. This is a standard technique and its application to the spin Hall effect is detailed in Ref.\ [\onlinecite{Our1}].

Diffusive transport in the absence of a external magnetic field is described by charge and spin current  \cite{Our1}
\begin{eqnarray}
\frac{e  }{\sigma}  {\bm{j}}_{\text{c}} \! \!  & =  \! \! & - \bm{\nabla }\mu _{\text c} +\frac{\alpha}{lk_F} \bm{\nabla }\times \bm{\mu }_{\text{s}} \label{10} \, , \\
\frac{e }{\sigma} \hat {\bm{j}}_{\text{s}} \! \!  & =  \! \! & - \bm{\nabla }( \hat{\bm{\sigma }} \bm{\mu }_{\text {s}} ) \! + \! \frac{\alpha }{l k_F} \hat{\bm{\sigma }}\times \bm{\nabla }\mu _c \! - \! \frac{2\alpha }{3}\hat{\bm{\sigma }} \times \bm{\nabla } \times \bm{\mu }_{\text{s}}
\label{20} \, ,
\end{eqnarray}
where $\sigma $ is the conductivity, $\alpha $ is a dimensionless spin-orbit coupling constant, $k_F$ is the Fermi wavevector, $l$ is the mean free path. $\mu _{\text c},\bm{\mu }_{\text s}$ are charge and spin 
accumulations, and $\hat{\bm{\sigma }}$ is the Pauli matrix, so that the charge density and spin density are $n=N_0\mu _c$ and $\bm{s}=N_0\bm {\mu _{\text s}}$, where $N_0$ is the density of states. The first and second terms in (\ref{10}) are the ordinary current and the anomalous current due to the anomalous spin-orbit velocity operator, respectively. The first and second terms in (\ref{20}) have a similar origin. The last term in the spin current (\ref{20}) is due to side-jump {\it and} skew scattering induced by the spin-orbit interaction. In our notation, the $2 \times 2$ spin current (\ref{20}) decomposes as $\hat {\bm{j}}_{\text{s}}={\bm j}_{\text{s}}^x\hat \sigma _x+{\bm j}_{\text{s}}^y\hat \sigma _y+{\bm j}_{\text{s}}^z\hat \sigma _z$, where the vectors ${\bm j}_{\text{s}}^x$, ${\bm j}_{\text{s}}^y$ and ${\bm j}_{\text{s}}^z$ represent flow of particles with spins along $x$, $y$, and $z$, respectively.  

Introducing a magnetic field gives an additional contribution to the electron self-energy, the Zeeman energy, associated with the coupling between the magnetic field and the spin of the electrons. We assume the Zeeman energy is small compared to the elastic scattering rate, $b \equiv g_L\mu _B|\bm{B}|\tau /\hbar \ll 1$, where $\mu _B$ is the Bohr magneton, $g_L$ is the gyromagnetic ratio and $\bm{B}$ is 
the external magnetic field. We also disregard higher order contributions of the order $\alpha b/(k_Fl)$. Additional spin current governed by the external magnetic field is in this regime:
\begin{equation}
\frac{e}{\sigma} \hat{\bm j}_{\text s}^B=\frac{g_L\mu _B\tau}{\hbar }\bm{\nabla }(\bm{B}\cdot \bm{\mu }_{\text s}\times \hat{\bm{\sigma }}) \label{70} 
\end{equation}
In general, the orbital Lorentz force should also be included. It gives additional contributions to the charge current ${\bm j}_{\text c}^{L}= (\sigma \tau/m) \bm{B} \times \bm{\nabla }\mu _{\text c}$ and spin current $\hat{\bm j}_{\text s}^{L}=( \sigma \tau/ m)(\bm{B}\times \bm{\nabla })(\bm{\mu }_{\text s}\hat{\bm{\sigma }})$. For the thin film geometry with in-plane magnetic field discussed below, the Lorentz force vanishes, and we will consequently not dicuss the orbital contribution in the remainder of our work.

We assume the spin-orbit scattering is stronger than the spin-flip scattering due to magnetic impurities. The diffusion equations for the charge and spin-dependent potentials are then 
\begin{eqnarray}
\bm{\nabla ^2}\mu _c & = &0 \, , \label{40} \\
{\bm{\nabla ^2\mu}_{\text s}} & = & \frac{\bm \mu_{\text{s}}}{l_{\text{sf}}^2}+\frac{\bm{n}_B\times \bm{\mu_{\text s}}}{l_{\text{m}}^2}-\frac{\bm{n}_B\times \bm{\mu_{\text s}}\times \bm{n}_B}{l_{\text{sm}}^2}  \, , \label{50} 
\end{eqnarray}
where $\bm{n}_B$ is a unit vector along $\bm{B}=B \bm{n}_B$. 
As it is well known, spins decay on the spin-orbit induced length scale $l_{\text {sf}}=\sqrt{D\tau _{\text {so}}}$ and precesses around the magnetic field on the length scale $l_{\text {m}}=\sqrt{\hbar D/(g_L\mu _BB)}$, where $D=v_F^2\tau/3$ and $\tau _{\text{so}}$ is the spin-flip relaxation time due to the spin-orbit interaction, $1/\tau _{\text{so}}=8\alpha ^2/9\tau $ \cite{Our1}. The spin-precession length scale $l_{\text {m}}$ is comparable to $l_{\text {sf}}$ when $b\sim \alpha $. The last term in (\ref{50}) can be a source of spins competing with the spin-orbit induced spin relaxation and depends on the spin-magnetic length scale $l_{\text{sm}}=v_F\hbar /(\sqrt{3}g_L\mu _BB)$.\cite{zhang2} It is only important for large magnetic fields in Al, e.g. for the parameters discussed below when $B>10$ T, when $l_{\text{sm}}$ becames comparable to $l_{\text{sf}}$.  It becomes important at a lower magnetic field in Cu. The influence of the Zeeman-induced spin current (\ref{70}) is also small at fields below $B\sim 10$ T in Al. We employ below this theory to calculate transverse spin and charge accumulations in a thin normal metal film.

\begin{figure}[ht]
\includegraphics[angle=0,width=7cm]{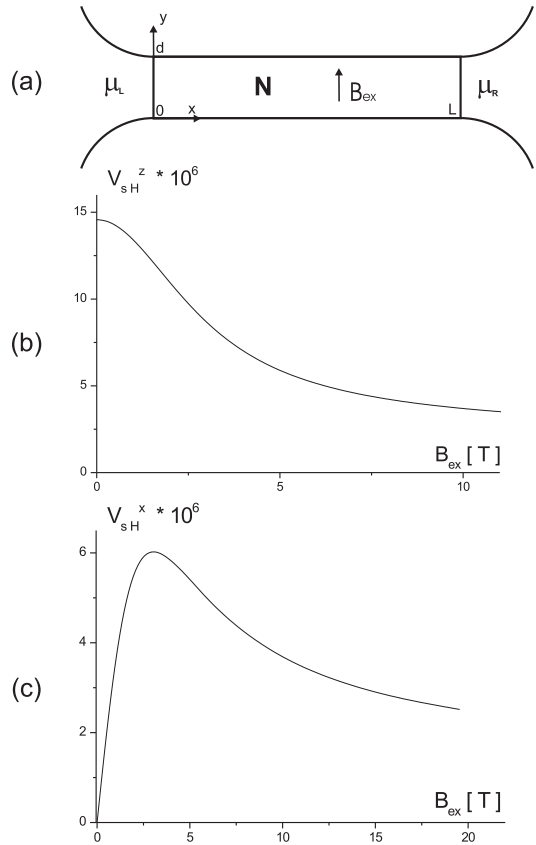}
\caption{Thin metallic film with contacts to reservoirs ({\bf a}). Relative spin Hall accumulations in the middle of the film ($x=L/2$) as a function of external magnetic field for $z$-spin component ({\bf b}) and for $x$-spin component ({\bf c}). The field direction $\bm{B_{\text{ex}}}=(0,B_{\text{ex}},0)$, sample sizes are $d=282\,nm, L=1100\,nm$.}
\label{fig:normal}
\end{figure}

\begin{figure}[ht]
\includegraphics[angle=0,width=7cm]{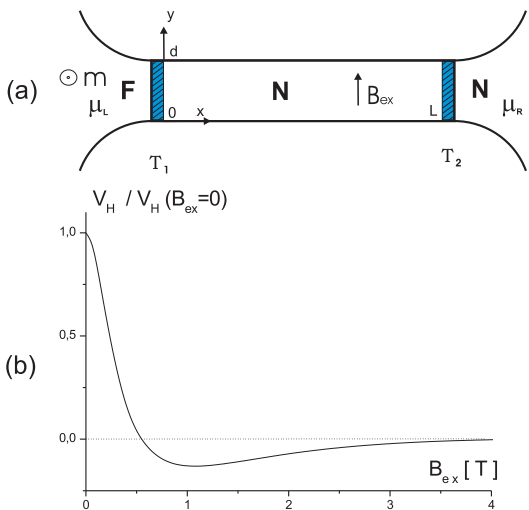}
\caption{Thin metallic film with tunneling contacts $T_1$ and $T_2$ to ferromagnet with magnetization $\bm{m}=(0,0,1)$ in the left side and with a normal metal in the right side ({\bf a}). The normalized relative Hall accumulation in the middle of the film ($x=L/2$) as a function of external magnetic field ({\bf b}). The field direction $\bm{B_{\text{ex}}}=(0,B_{\text{ex}},0)$, sample sizes are $d=282\,nm, L=1100\,nm$.}
\label{fig:SHx_NorMet}
\end{figure}

We consider two configurations where spin Hall and Hall potentials can be measured: when the normal metal film is attached to a) normal metal reservoirs via metallic contacts and b) ferromagnetic reservoirs via tunnel contacts, as shown in Fig.\ \ref{fig:normal}a and Fig.\ \ref{fig:SHx_NorMet}a correspondingly. In the first case, no spins are injected, but a transverse spin accumulation (spin Hall effect) builds up due to the spin-orbit interaction in zero magnetic field. This spin accumulation can \textit{e.g.} be measured via the charge potentials in ferromagnetic reservoirs connected to the transverse edges via tunnel contacts. In the second case, spins are injected, and the system exhibits a transverse spin-orbit induced charge accumulation at zero magnetic field. This charge potentials can be measured via the charge potentials in normal metal probes.

 The diffusion equations (\ref{40}) and (\ref{50}) are solved numerically on a sufficiently fine grid giving convergent results with the proper boundary conditions for incoming charge and spin flow through the contacts to the reservoirs. The diffusion equation also must be supplemented by a transverse boundary condition: There is no particle or spin flow across the transverse boundaries at $y=0$ and $y=d$.

Let us first consider scenario a) when a thin normal metal film of length $L$ and width $d$ is attached with perfect contacts with zero resistance to a left reservoir with local chemical potential $\mu_{\text{L}}$ and a right reservoir with local chemical potential $\mu_{\text{R}}$, as shown in Fig.\ \ref{fig:normal}a. 
We bias the system along the $x$ direction by a symmetric potential $\mu _{\text L}=-\triangle \mu /2$ and $\mu_{\text R}=\triangle \mu /2$ and consider the current and density response. We consider a transverse in-plane magnetic field, directed along $y$.

We use values for the spin-diffusion length and spin-flip relaxation time in Al at low temperature\cite{Jedema}, $l_{\text{sf}} = 600$ nm, $\tau _{\text{so}}=90$ ps, and  $v_F=2.03\cdot 10^{6}\,{\text m}\,{\text s}^{-1}$ (ref. [\onlinecite{Ashcroft}]), e.g. $\alpha =0.006, \tau =2.9 \cdot 10^{-15}\,{\text s}, l=5.9\,nm$. The sample width $d=282$ nm and length $L=1100$ nm are assumed. 

The charge Hall effect is small in this case since the Lorentz force vanishes.  An out-of-plane spin Hall potential, \textit{e.g.} a spin accumulation along the $z$ axis, is induced by the anomolous velocity operator\cite{zhang,Our1} and it is affected by spin-precession in a magnetic field (\ref{50}). The most interesting is the in-plane spin Hall effect, where the spin accumulation is directed along $x$. This component vanishes at zero magnetic field\cite{zhang,Our1}. It increases at weak magnetic fields due to the precession of spins initially directed along $z$. Thus, there is an in-flux of spins from the $z$ component (induced by the spin Hall effect) to the $x$-component.  However, the spin accumulation start to decrease for large fields due to decoherence caused by precession and spin relaxation. We are consequently interested in the dependence of the $x$- and $z$-spin Hall effects on an external magnetic field. To this end, we introduce the corresponding relative spin and charge accumulations as $V_{\text{sH}}^i(x)=[\mu _{\text s,i}(x,0)-\mu _{\text s,i}(x,d)]/\triangle \mu$ and $V_{\text{H}}(x)=[\mu _{\text c}(x,0)-\mu _{\text c}(x,d)]/\triangle \mu $, where $i = x, y, z$. There is also a spin Hall accumulation along the $y$ direction (not shown).

Numerical results are presented in Fig. \ref{fig:normal}. 
We see that the out-of-plane spin Hall effect, $V_{\text{sH}}^z$, is suppressed by the magnetic field (see Fig. \ref{fig:normal}b). This occurs because there is no real spin-injection into the system, but the spin polarization arises within the sample due to the spin-orbit interaction. Consequently, precession takes places from many origins within the sample and the average of all initial conditions results in a decaying out-of-plane spin Hall effect as a function of magnetic field. In our normal metal geometry, the decay of the out-of-plane spin Hall effect is not exactly Lorentzian as in the semiconductor case,\cite{Exper} presumably because the physics is different for semiconductors and normal metals with different effective masses, Lande g-factors, Fermi energies and microscopic details of the spin-orbit interaction. The dependence of the in-plane spin Hall effect is qualitatively different from that of the out-of-plane spin Hall effect. As discussed above, the numerical results in Fig. \ref{fig:normal}c demonstrated that there is a critical field $B_{\text{ex}}\approx 2.93\,\text{T}$ where $V_{\text{sH}}^x$ attains its maximum. On the left of the maximum the $z$-spins precession creates a spin accumulation along $x$ which was initially absent at zero magnetic field. The spin accumulation increases with increasing magnetic field $B$. In the right side from the maximum the situation is qualitatively different, and we observe a decay of the accumulation because $x$- and $z$-spins start to relaxate due to strong spin precession. In the critical point the ratio of the magnetic length scale to spin-orbit length scale are similar, $l_{\text m}/l_{\text{sf}}\simeq 0.2$.

Next, we consider geometry b) when a normal metal is in tunnel contact to ferromagnet and normal metal reservoirs, see Fig.\ \ref{fig:SHx_NorMet}a. We assume that also in this case an in-plane, $y$-directed, magnetic field is applied, and that the ferromagnet magnetization is out-of-plane, e.g. directed along $z$. To describe the transport properties, we 
introduce the conductance of the normal metal $G_0=\sigma wd/L$, where $w$ is the width of the film. The conductance at the tunnel barrier between the normal metal film and the normal metal reservoir is $G$. The contact between the ferromagnetic 
reservoir and the normal metal film is described by the spin-dependent conductances $G^{\uparrow}$ and $G^{\downarrow}$ for spin aligned and antialigned to the magnetization and a mixing conductance of reflection $G^{\uparrow \downarrow}$ for spins in 
the normal metal that are non-collinear to the magnetization direction.\cite{Brataas:prl00}

In order to determine the spin and charge accumulations, we need the boundary conditions for the spin and charge flow through the tunnel contacts from the ferromagnetic reservoir into the normal metal and from the normal metal into the normal metal reservoir. These boundary conditions are determined by magnetoelectronic circuit theory\cite{Brataas:prl00,Our1}. 
On the left boundary, the charge current along the transport, $x$, direction through the tunneling contact $T_1$ can be written as
\begin{equation}\label{160}
e I_c=(G^{\uparrow }+G^{\downarrow })\left[\mu_c^F-\mu_c(0,y)\right]-(G^{\uparrow }-G^{\downarrow })\bm{m}\cdot \bm{\mu}_{\text{s}}(0,y)
\end{equation}
in terms of the local chemical potential in the ferromagnet $\mu_c^F\equiv \mu_L$ and the  spin and charge chemical potentials in the normal metal close to the ferromagnetic interface $\bm{\mu}_{\text{s}}(x=0,y)$ and $\mu_{\text{c}}(x=0,y)$. 
Here $\bm{m}$ is unit vector in the direction of the magnetization in the ferromagnet. For a tunnel contact $2{\text{Re}}G^{\uparrow \downarrow }=G^{\uparrow }+G^{\downarrow }$. The spin-current along the transport, $x$, direction is then
\begin{equation}\label{170}
e\bm{I}_{\text{s}}=\bm{m}(G^{\uparrow }-G^{\downarrow })(\mu_c^F-\mu_c(0,y))-2{\text{Re}}G^{\uparrow \downarrow }\bm{\mu}_{\text{s}}(0,y).
\end{equation}
Similarly, the spin and charge current can be found on the normal metal side of the junctions from the diffusion equations. The boundary condition is determined by continuity of spin and charge currents close to the tunnel barrier. Similarly, we can find the boundary condition for the right tunnel barrier in contact with the normal metal reservoir where there is no spin-dependence of the contact conductance. In discussing our results, we introduce the the relative conductance $q=G_{\text o}/G$, the tunnel barrier polarization $p=(G^{\uparrow }-G^{\downarrow })/(G^{\uparrow }+G^{\downarrow })$. We use reasonable values $G=G^{\uparrow }+G^{\downarrow }$, $p=1/2$, and  $q=1$.

In this geometry, the spin-orbit interaction converts the incoming spin flux via the anomalous current operator into a transverse charge Hall voltage even in the absence of a magnetic field. The magnitude of this charge Hall accumulation is approximately one order of magnitude smaller than the $z$-spin Hall accumulation in geometry a) at zero field, $V_H(x=0, B_{\text{ex}}=0)\simeq 1.5\cdot 10^{-6}$. Numerical results for the relative Hall voltage as a function of magnetic field $B_{\text{ex}}$ are presented in Fig. \ref{fig:SHx_NorMet}b. When we introduce the external magnetic field, the spins initially directed along $z$ starts precessing. The spins are induced within the sample. Precession from many spatial origins results in a full destruction of the Hall effect when $B \sim 3 \div 4\,{\text{T}}$, which is close to the critical field for the in-plance ($x$) spin Hall effect (see Fig.\ \ref{fig:normal}c). For smaller magnetic fields, precession can lead to a reversal of the charge Hall effect. The magnetic field range in which we can observe precession via the charge Hall potential is naturally similar to what is observed in Ref.\ \onlinecite{Jedema} on spin precession in Al.

In summary, we have studied transport in normal metals with small spin-orbit interaction in a external magnetic field. Based on analytical diffusion equations for charge and spin potentials, numerical results for spin Hall and charge Hall potentials in experimentally relevant geometries are obtained. We studied thin film normal metals where a) no spins are injected from normal metal reservoirs and b) spins are injected via ferromagnetic reservoirs. Spin precession destroys the out-plane spin Hall potential when the Larmor frequency is larger than the spin-orbit induced spin relaxation length. It similarly, also destroys the charge Hall potentials which can be induced when spins are injected into the normal metal films. The in-plane spin Hall effect initially vanishes at zero magnetic field and then increases for weak magnetic fields due to spin precession from the initial out-of-plane spin Hall accumulation. For larger magnetic fields, the in-plane spin Hall effect also decays. 

This work has been supported in part by the Research Council of Norway, NANOMAT Grants No. 158518/143 and 158547/431, and through Grant No. 153458/432.


\begin{thebibliography}{99}
%
\bibitem{dyak1} M. I. Dyakonov and V. I. Perel, Sov. Phys. JETP {\bf 33}, 467 (1971).
\bibitem{dyak2} M. I. Dyakonov and V. I. Perel, Phys. Lett. A {\bf 35}, 459 (1971).
\bibitem{chaz} J. N. Chazalviel, Phys. Rev. B {\bf 11}, 3918 (1975).
\bibitem{hirsh} J. E. Hirsch, Phys. Rev. Lett. {\bf 83}, 1834 (1999).
\bibitem{Exper} Y. K. Kato, R.S. Myers, A.C. Gossard, D.D. Awschalom, Science {\bf 306}, 1910 (2004).
\bibitem{murakami} S. Murakami, N. Nagaosa and S.-C. Zhang, Science {\bf 301}, 1348 (2004).
\bibitem{sharma} P. Sharma, Science {\bf 307}, 531 (2005).
\bibitem{wolf} S.A. Wolf, D.D. Awschalom (Eds.), \textit{Spintronics: A Spin-Based Electronics Vision for the Future},
 Science {\bf 294}, 1488 (2001)
\bibitem{ganichev} S.D. Ganichev, E.L. Ivchenko (Eds.), Nature {\bf 417}, 153 (2002). 
\bibitem{sinitsyn} N. A. Sinitsyn, E. M. Hankiewicz, Winfried Teizer, and Jairo Sinova, Phys. Rev. B {\bf 70}, 081312 (2004)  
\bibitem{culcer} D. Culcer, J. Sinova, N.A. Sinitsyn (Eds.), Phys. Rev. Lett. {\bf 93}, 046602 (2004). 
\bibitem{burkov} A.A. Burkov, A.S. Nunez and A. H. MacDonald, Phys. Rev. B {\bf 70}, 155308 (2004).
\bibitem{new1} N. Sugimoto, S. Onoda, S. Murakami, and N. Nagaosa, cond-mat/0503475.
\bibitem{Jedema} F. J. Jedema, H. B. Heersche, A. T. Filip, J. J. A. Baselmans and B. J. van Wees, Nature {\bf 416}, 713 (2002).
\bibitem{chang} M.-C. Chang, cond-mat/0411697.
\bibitem{shen} S.-Q. Shen, M. Ma, X.C. Xie, and F.C. Zhang, Phys. Rev. Lett. {\bf 92}(25), 256603 (2004). 
\bibitem{sinova} J. Sinova, D. Culcer, Q. Niu, N. A. Sinitsyn, T. Jungwirth, and A. H. MacDonald, Phys. Rev. Lett. {\bf 92}, 126603 (2004).
\bibitem{dyak} M. I. Dyakonov and V. I. Perel, Zh. Eksp. Ter. Fiz. {\bf 13}, 657 (1971) [JETP {\bf 33}, 467 (1971)]
\bibitem{khae} A. Khaetskii, cond-mat/0408136 v1.
\bibitem{zhang} S. Zhang, Phys. Rev. Lett. {\bf 85}, 393 (2000).
\bibitem{hanle} G. Moruzzi and F. Strumia (Eds.), \textit{The Hanle effect and Level-Crossing Spectroscopy}, (New York: Plenum Press, 1991). 
\bibitem{schwab} P.~Schwab and R.~Raimondi, Ann. Phys. (Leipzig) 12, 471 (2003).
\bibitem{rammer2} J. Rammer, H. Smith, Rev. of Mod. Phys. Vol. {\bf 58}, No {\bf 2}, 323 (1986).
\bibitem{Our1} R.V. Shchelushkin and Arne Brataas, Phys. Rev. B {\bf 71}, 045123 (2005).
\bibitem{zhang2} Y. Qi and S. Zhang, Phys. Rev. B {\bf 67}, 052407 (2003).
\bibitem{Ashcroft} N. W. Ashcroft and N. D. Mermin, \textit{Solid State Physics} Ch.2 (ed. D. G. Crane) 38 (W.B. Saunders, Orlando, 1976). 
\bibitem{Brataas:prl00} A. Brataas, Yu. V. Nazarov, and G. E. W. Bauer, Phys. Rev. Lett. {\bf 84}, 2481 (2000).
%
\end{thebibliography}
\end{document}